\documentclass[manuscript]{aastex}
\begin{document}
\title{SPECTROSCOPIC STUDY OF IRAS\,19285+0517(PDS\,100): A RAPIDLY ROTATING Li-RICH K GIANT}
\author{Bacham E. Reddy and David L. Lambert}
\affil{Department of Astronomy, University of Texas, Austin, TX 78712; ereddy@astro.as.utexas.edu, dll@astro.as.utexas.edu}

\author{Bruce J. Hrivnak\altaffilmark{1}}
\affil{Department of Physics and Astronomy, Valparaiso University, Valparaiso, IN 46383; bruce.hrivnak@valpo.edu}
\altaffiltext{1}{
Visiting astronomer, Kitt Peak National Observatory.
KPNO is operated by the
Association of Universities for Research in Astronomy, Inc.\
under contract with the National Science Foundation.}

\author{Eric J. Bakker}
\affil{Leiden Observatory, P.O. Box 96864, 9513, 2300 RA Leiden, The Netherlands; bakker@strw.leidenuniv.nl}
\begin{abstract}

We report on photometry and high-resolution spectroscopy
for IRAS\,19285+0517. The spectral energy distribution based on
visible and near-IR photometry and far-IR fluxes shows
that the star is surrounded by dust at a temperature of $T_{\rm {d}}$ $\sim$ 250~K.
Spectral line analysis shows that the star is a K giant
with a projected rotational velocity $v\,sin i$ = 9 $\pm$ 2 km s$^{-1}$. 
We determined the atmospheric parameters: $T_{\rm {eff}}$ = 4500 K, log $g$ = 2.5, $\xi_{t}$ = 1.5 km s$^{-1}$, and
[Fe/H] = 0.14 dex.
The LTE abundance analysis shows that the star
is Li-rich (log $\epsilon$(Li) = 2.5$\pm$0.15), but with essentially
normal C, N, and O, and metal abundances.
Spectral synthesis of molecular CN lines yields the carbon isotopic ratio $^{12}$C/$^{13}$C = 9 $\pm$3, 
a signature of post-main sequence evolution and dredge-up on the RGB. 
Analysis of the Li resonance line at 6707~\AA\ for different ratios $^{6}$Li/$^{7}$Li
shows that the Li profile can be fitted best with a predicted profile for pure $^{7}$Li.
Far-IR excess, large Li abundance, and rapid rotation suggest
that a planet has been swallowed or, perhaps, that an 
instability in the RGB outer layers triggered a sudden
enrichment of Li and caused mass-loss.
\end{abstract}

\section{INTRODUCTION}
Observations of lithium in the atmospheres of red giants continue to
pose theoretical challenges. Stars evolving up the red giant branch
from the main sequence develop a deep convective envelope that
dilutes the surface lithium abundance. Thus, red giants are
predicted to have a lithium abundance considerably below the
initial or interstellar abundance of log $\epsilon$(Li)$\approx$3.0. Classical calculations
(Iben 1967) predict a reduction by a factor of 
30 and 60 for solar metallicity stars of 1~$M_\odot$ and 3~$M_\odot$,
respectively.
If some lithium has been destroyed
prior to evolution off the main sequence, the red giant's lithium
abundance will be even further reduced. However, as a very luminous asymptotic
red giant (4-7 M$\odot$), theory and observations show that surface lithium
can be replenished and even increased above the initial or interstellar
abundance thanks to hot bottom burning in intermediate-mass stars (Sackmann \& Boothroyd 1992).

This paper is concerned with a red giant that has yet to evolve to the
asymptotic giant branch but that has a lithium abundance greatly in
excess of that expected of a star on the red giant branch. Wallerstein \&
Sneden (1982) discovered the first such red giant: HD\,112127 with
$\log \epsilon$(Li) $\simeq$ 3.0, i.e., the interstellar value. In the last
two decades, additional examples have been discovered; Charbonnel \&
Balachandran (2000) list 17 stars (including 3 subgiants, 2 early-AGB stars, and
2 for which evolutionary status has yet to be determined) with  $\log \epsilon$(Li) $\geq$ 2.0,
and include several additional examples with a lower lithium abundance which is most
probably in excess of that expected of a red giant.
Three of the 17 stars have a lithium abundance clearly greater than
the stars' probable initial value. Charbonnel \& Balachandran (2000)
lists $v\, sin i$ values for 22 giants of which 10 have $v\, sin i$ $>$ 8 km s$^{-1}$.
If we include the PDS~365 (Drake et al. 2001) and the IRAS~19285+0517 (present study),
the number of Li-rich giants for which $v\, sin i$ $>$ 8 km s$^{-1}$ rises to 12.
It now appears that more than half of the Li-rich giants are rotating at a unusually
high rate for normal K giants for which average $v\,sin i \sim$ 2.0 (de Medeiros et al. 2000). 
This proportion drops to a mere 2\% for the more common
slowly rotating K giants (Drake et al. 2001). 
Most of them
exhibit a
pronounced infrared excess (Gregorio-Hetem et al. 1992; Gregorio-Hetem,
Castilho, \& Barbuy 1993). 

Here, we report an analysis of a K giant with large infrared
excess, first reported to be a Li-rich candidate by de la Reza et al. (1997).
The star is IRAS 19285+0517, also known as PDS100 after the
Pico dos Dias Survey (de la Reza, Drake, \& da Silva 1996). No prior
quantitative spectroscopy of this star has been reported.
We show from high-resolution echelle spectra that the star is Li-rich
($\log \epsilon$(Li) $\simeq$ 2.5) and rapidly rotating ($v\,sin i$ $\simeq$
9 km s$^{-1}$). 

\section{OBSERVATIONS}
\subsection{Photometry}

Standardized photometry was carried out for this object at the Kitt Peak
National Observatory (KPNO).
Visible photometry was made using a CCD on the
0.9 m telescope on 1995 September 13 and near-infrared observations were made
using the Cryogenic Optical Bench (COB) infrared camera on the 2.1 m
telescope on 1995 October 12.  The results are listed in Table 1.

The object is red, with B$-$V = 1.48.
Some of the reddening may be due to interstellar extinction and some also
due to circumstellar extinction. An effective reddening of $E(B-V) = (B-V)_{obs} - (B-V)_{pred} = 0.34$ was
estimated using a predicted color B-V = 1.14 for the
derived atmospheric model (next section) and synthetic colors 
computed by Girardi et al. (2000). The visible photometry is
corrected for extinction using the extinction curve computed by Seaton (1979), and
the near-infrared values are corrected for extinction using
the relationship between infrared and visible extinction (Cardelli, Clayton, \& Mathis 1989).
Flux-converted photometry
is compared with the Kurucz (http://cfaku5.harvard.edu) 
flux model for $T_{\rm {eff}}$ = 4500~K, log $g$ = 2.5, [Fe/H] = 0.0 in Figure~1.
IRAS infrared fluxes
at 12~$\mu$m, 25~$\mu$m, and 60~$\mu$m are well above the expected
fluxes, indicating
an infrared excess. A black body fit to the IRAS fluxes
indicates a dust temperature ($T_{\rm d}$) of about 250~K. In sharp contrast, the
IRAS fluxes of the K giant Arcturus are well fit by the predicted 
photospheric fluxes (see Figure~1).
\subsection{High-Resolution Spectroscopy}

A high-resolution  spectrum of IRAS\,19825+0517 was obtained on
1997 October 17 with the McDonald Observatory's 2.7 m telescope using the
cross-dispersed echelle spectrograph (Tull et al. 1995). Three spectra
were recorded on a CCD, each with an exposure of 20 minutes. A Th-Ar hollow
cathode lamp was observed between the stellar exposures. The spectra
were reduced in the standard fashion using the IRAF\footnote{IRAF is distributed
by the national Optical Astronomical Observatories, which is operated
by the Association for Universities for Research in Astronomy, Inc.,
under contract to the National Science Foundation.}. 
The final combined spectrum has a signal-to-noise
ratio S/N $\simeq 350$ in the continuum at 6500 \AA\ and
a resolving power of $\lambda/\Delta\lambda \simeq 55,000$, as measured from 
the comparison lines. The usable spectrum covers the wavelength interval
from about 5000 \AA\ to 9500 \AA\ with gaps between echelle orders longward
of 5680 \AA. Below 5000 \AA\ the lines are too blended and the S/N ratio is too low for analysis.

Inspection of the spectrum confirmed the great strength of the Li\,{\sc i} 6707~\AA\ line, the 
atypical widths of all absorption lines, and unusual profiles of the Na D and H$\alpha$ lines (Figure~2).
Both H$\alpha$ and Na~D
profiles are seen to be asymmetric.
An asymmetric H$\alpha$ profile is common in Li-rich K giants (Drake et al. 2001) and is
attributed to chromospheric activity in the star. The core (deepest part of the line) of the H${\alpha}$ 
profile is at the systemic velocity (V$_{r}$) of 2.0$\pm$1.5 km s$^{-1}$. 
The Na~D profiles are complex. The line cores are asymmetric like H$\alpha$.
There is a strong absorption at $-$16 km s$^{-1}$ with
respect to the systemic velocity. This may be due to circumstellar gas but
we can not exclude the possibility of an interstellar origin. More
interesting is the shallow broad blue-shifted absorption. 
This component is surely of circumstellar origin.

\section{ANALYSIS}
For the abundance analysis we adopted the widely used grid of LTE model atmospheres computed
by Kurucz (http://cfaku5.harvard.edu). The models are available in steps of 250~K ($T_{\rm eff}$),
0.5~dex (log $g$), and 0.5~dex ([M/H]). Required models were interpolated.
Another basic requirement in the abundance analysis is a set of reliable
$gf$-values for the selected lines. 
For many of the transitions, laboratory measured $gf$-values are available.
For the rest of the transitions we chose reliable theoretical values.
For iron, a
large number of Fe\,{\sc i} and Fe\,{\sc ii} line $gf$-values were reviewed by
Lambert et al. (1996).
For other elements, we used the large compilation of R.E. Luck (1993, private communication).
All of the atomic data were examined by deriving abundances for the Sun and the K giant Arcturus.
Equivalent widths ($W_{\lambda}$) for the Sun were measured using
the spectrum of the asteroid Iris (reflected solar spectrum) 
obtained with the same instrument as that used for our program star.
In the case of Arcturus, we used high-resolution digitized spectra published
by Hinkle et al. (2000).

\subsection{Atmospheric Parameters}

Effective temperature ($T_{\rm eff}$), surface gravity (log $g$),
microturbulent velocity ($\xi_t$), and chemical composition (in short,
[Fe/H]) are essential atmospheric parameters. For IRAS\,19285+0517, 
the atmospheric parameters are determined spectroscopically. 
Standard procedures are not
directly applicable here
because the high rotational velocity increases the blending of lines
and reduces the number of useful lines.  In a modification of the standard
procedure, we first determined $\xi_t$
and then proceed to the determination of the latter
parameters.

To determine $\xi_t$ we chose combination of weak and strong lines of the same
species and with similar lower excitation potentials (LEP). Differences
 in abundance from the strong and weak line (DABUN(s-w)) were
computed for a range of $\xi_t$. Line combinations from Fe\,{\sc i}, Co\,{\sc i},
and Ni\,{\sc i} were considered; details are given in Table 2.
The condition DABUN(s-w) = 0 provides
$\xi_t$. Figure 3 shows the run of DABUN(s-w) vs $\xi_t$ for the
selected pairs. All intersect DABUN(s-w) = 0 at a similar value of
$\xi_t$; we adopt $\xi_t$ = 1.5 $\pm$ 0.1 km s$^{-1}$. This
result is insensitive to the adopted $T_{\rm eff}$ and $\log g$. 
We found 
the $T_{\rm eff}$ insensitivity
for  pairs of Fe\,{\sc i}, Co\,{\sc i}, and Ni\,{\sc i} lines by computing DABUN(s-w)
for four different values of $\xi_t$ for models spanning 600 K in
$T_{\rm eff}$. 

Having determined $\xi_t$, we found $T_{\rm eff}$ in the usual way by
demanding that a model atmosphere return the same Fe (Ni) abundance
from a large number of Fe\,{\sc i} (Ni\,{\sc i})
 lines with a range in LEP. The 38 Fe\,{\sc i} lines used
range in LEP from 0.96 eV  to 5.0 eV and the 32 Ni\,{\sc i} lines
range from 1.6~eV to 4.4~eV. Surface gravity has little effect
on the $T_{\rm eff}$ determination.
Both sets of lines yield $T_{\rm eff} = 4500 \pm 100$ K.

Application of ionization balance using the previously determined
$\xi_t$ and $T_{\rm {eff}}$ then gives an estimate of the surface gravity with 
the abundance of the metal used in the exercise.
Neutral and singly-ionized lines of Fe, Cr, and Ti give three
independent estimates of $\log g$. The number of lines in each solution
is ($n_{atom}, n_{ion})$ = (32,8), (8,2), and (9,4), and these lead to
$\log g$ solutions of 2.7, 2.3, and 2.5 for
Fe, Cr, and Ti, respectively. We adopt a mean value $\log g$ =
2.5 $\pm$ 0.25. The iron abundance (relative to the Sun) determined is [Fe/H] = +0.14$\pm0.16$.

As a check on our choice of lines and atomic data, we derived atmospheric
parameters for Arcturus, a slowly rotating bright K giant of similar $T_{\rm {eff}}$ and log $g$ but
of lower [Fe/H].  
For our selection of lines, we measured
equivalent widths ($W_{\lambda}$) from the high-resolution spectral atlas
published by Hinkle et al. (2000). A determination of $\xi_t$
from the pairs of Fe\,{\sc i} and Ni\,{\sc i} lines gave
$\xi_t = 1.6 \pm 0.3$ km s$^{-1}$. The Fe\,{\sc i} lines of different LEP
required a model with $T_{\rm eff} = 4275 \pm 50$ K. Ionization
balance applied to Fe\,{\sc i} and Fe\,{\sc ii} lines gave $\log g =
1.5 \pm 0.25$ with [Fe/H] =  $-$0.58 $\pm$ 0.1.
The results from these selected lines are in 
excellent agreement with the results of Peterson et al. (1993) from a very
detailed analysis, namely $T_{\rm eff} = 4300 \pm 30$ K,
$\log g = 1.5 \pm 0.15$, $\xi_{\rm {t}} = 1.7$ km s$^{-1}$, and [Fe/H] = $-$0.50 $\pm$
0.1. 
The derived atmospheric values
are given in Table~3.
\subsection{Physical Parameters: $v\,sin i$, V$_{r}$, and L}
The projected rotational velocity of IRAS\,19285+0517 was estimated by fitting a
synthetic spectrum to the portions of the observed spectrum. 
In addition to microturbulence, collisional broadening, and rotation,
we included macroturbulence. This was modeled as a Gaussian distribution
function with parameter $V_m$ for which we
adopted $V_m = 3$ km s$^{-1}$, a value typical of K giants (Fekel 1997). 
Using the current version of the spectrum synthesis code MOOG (Sneden
1973), we synthesized profiles of a few strong  lines. The standard
form of rotational broadening was assumed (Gray 1992) with a limb-darkening
coefficient $\epsilon$ taken from Wade \& Rucinski (1985). 
Figure 4 shows fits to Fe\,{\sc i} and Ni\,{\sc i}
lines at 6173.3~\AA\ and 6175.3~\AA, respectively.
Observed profiles are fitted by varying only $v\,sin i$ and 
keeping
macroturbulence, instrumental profile width (FWHM = 0.15~\AA), 
and abundances constant.
Note that the predicted profile for the $v\,sin i$ = 2.0 km s$^{-1}$
for normal K giants (De Medeiros et al. 2000) 
is far too sharp to match the observed profile.
A profile for
$v\,sin i = 9 \pm 2$ km s$^{-1}$ best fits the observed profiles.
The result is insensitive to the adopted value of $V_{\rm {m}}$;
if the macroturbulence is neglected,
$v\,sin i = 9.5 $ km s$^{-1}$ is found. 

Another important quantity is the heliocentric radial velocity $V_{\rm {r}}$ = 2.0$\pm$ 1.5 km s$^{-1}$
(Observing date: 1997 October 17 and 01:37:22 UT), derived
using the wavelength shifts of many symmetric absorption features.
The measurements of V$_{r}$ for lines of LEP in the range of 0 - 9 eV
show that the measured $V_{\rm {r}}$ is independent of LEP.
 
The evolutionary status of IRAS\,19285+0517 may be assessed from its luminosity,
which we determine from the standard expression derived from the relations
$L \propto R^2T_{\rm eff}^4$ and $g \propto M/R^2$:

\begin{equation}
\log \frac{L}{L_\odot} = \log \frac{M}{M_\odot} - \log g + 4\log T_{\rm eff} -10.61
\end{equation}

Given the derived $T_{\rm eff}$ and $\log g$, we obtain $\log L/L_\odot = 1.5\pm0.2$
for an assumed mass of 1$M_\odot$ and $\log L/L_\odot = 1.8$ for
2$M_\odot$. 
As judged by its location in the H-R diagram, IRAS\,19285+0517 is probably a He-core
burning or clump giant. Evolutionary tracks computed by Bertelli et al. (1994) for the
composition Z = 0.02 and Y = 0.28 put the clump giants at log L/L$_{\odot}$ = 1.7 and $T_{\rm {eff}}$ = 4500 K
after 10$^{10}$ yrs for a mass M $\approx$ 1 M$_{\odot}$ at the main sequence turn-off.

\section{CHEMICAL COMPOSITION}

A thorough abundance analysis was made using a Kurucz convective model
atmosphere having the adopted atmospheric parameters. For selected
elements, spectrum synthesis was conducted but for most elements unblended
lines could be found. We discuss the results under the headings: general
metal abundance, lithium, and  C, N, and O abundances.

\subsection{General Metal Abundances}

Lines of elements from Al to Nd were identified despite the moderate
rotational line broadening. Identifications were checked against the
high quality spectrum of Arcturus (Hinkle et al. 2000).
The entire line list was checked
against equivalent widths measured from the spectrum of the asteroid Iris,
and where necessary $gf$-values were adjusted. Table 4 summarizes the
abundances $\log\epsilon$(X) and the abundance ratios relative to solar abundances 
[X/H] and [X/Fe]. 
Solar abundances (log $\epsilon$(X)$_{\odot}$),
except for C, N, and O, are adopted from Grevesse \& Sauval (1998). For C, N, and O
we adopted the
abundances derived in this study using the solar spectrum. 
Uncertainties in the derived abundances are given in the form of $\sigma$$_{tot}$, where
$\sigma$$_{tot}$ is the quadratic sum of various sources of uncertainties: 
the $gf$-values, the measured W$_{\lambda}$ (represented by line-to-line scatter), and 
the derived atmospheric parameters (Table~3) . During the abundance analysis we estimated
the uncertainties in the model parameters: $\delta$$T_{\rm eff}$ = 100~K, 
$\delta$log $g$ = 0.25, 
and $\delta$$\xi_{\rm {t}}$ = 0.25 km s$^{-1}$.

To within the errors of measurement, abundances from Al to Nd are
normal for a star that is slightly metal-enriched relative to the
Sun, i.e., [X/Fe] $\simeq$ 0.0 is expected. Specifically, the $s$-process
elements are not enriched, confirming that the star is neither
a resident of the AGB nor has received mass from a companion that
had evolved to the AGB. 

\subsection{Lithium Abundance}

Initial inspection of the spectrum confirmed de la Reza et al.'s (1997)
identification of IRAS~19285+0517 as Li-rich. The Li\,{\sc i} 6707 \AA\
resonance doublet is very strong with an equivalent width
$W_\lambda \simeq 372$ m\AA. Owing to the blending of fine,
hyperfine, and possibly isotopic components, the doublet
must be analysed using spectrum synthesis. The wavelengths and line
strengths of these components are taken from Smith, Lambert, \& Nissen (1998),
as corrected by Hobbs, Thorburn, \& Rebull (1999). 

To complete the spectrum synthesis, a line list for several {\AA}ngstr\"{o}ms
around 6707 \AA\ was compiled, primarily from Kurucz (http://cfaku5.harvard.edu). We required that
this list reproduce the spectra of the Sun and  Arcturus, both Li-poor
objects. Then synthetic spectra for IRAS\,19285+0517 were computed and
broadened with the instrumental profile, the adopted macroturbulence,
and rotational broadening. A good fit to the Li\,{\sc i} line was
obtained with the abundance $\log\epsilon$(Li) = 2.5$\pm$0.1 (Figure~5, top panel) and the
assumption that $^6$Li is absent (see below).
At this Li abundance, the excited Li\,{\sc i} line at 6104 \AA\
is also expected to be present. Our synthesis of this weak line (Figure~5, bottom panel)  confirms the
abundance derived from the resonance doublet.
At $\log\epsilon$(Li) $\simeq$ 2.5, IRAS\,19285+0517
certainly enters the small but growing ranks of Li-rich red giants; the
Li abundance is about 1 dex greater than the maximum value expected of
a red giant with a deep convective envelope. 

The assumption that all lithium is $^7$Li deserves scrutiny. Addition of $^6$Li to the composition
increases the red asymmetry of the line and shifts it to longer
wavelengths (Figure~6). 
Synthetic profiles for different $^{6}$Li/$^{7}$Li ratios by varying
total lithium abundance were computed
such that the $W_{\lambda}$ of the predicted profile
matches with that of the observed profile ($W_{\lambda}$ = 370$\pm$3m\AA). 
The predicted profiles of equal $W_{\lambda}$, but for different $^{6}$Li/$^{7}$Li ratios and
Li abundances are compared with the overall shape of the observed profile
(Figure~6; upper panel). (The Li profile is significantly
saturated and addition of $^{6}$Li components require less total Li abundance (relative
to $^{7}$Li only profile)
to achieve the same $W_{\lambda}$ of $^{7}$Li only profile). 
The predicted profile of $^{6}$Li/$^{7}$Li = 0.0 and log $\epsilon$(Li) = 2.5$\pm$0.1
best fits with the observed profile, 
and ratios
greater than about 0.03 are excluded. 
Fitting has been done by fixing the observed
Li profile position (6708.832$\pm$0.02~\AA) determined from Doppler shifts of symmetric Fe\,{\sc i} and Ca\,{\sc i}
features for which accurate laboratory measurements are available. Within the measurement
errors observed Li profile position matches with the predicted profile 
position $\lambda_{c}$ = 6707.814~\AA\ (for $^{7}$Li only).
Lines with similar  sensitivity  as Li\,{\sc i} 6707 \AA\ to atmospheric 
structure are well reproduced by the synthetic spectra (e.g., the intercombination resonance
Ca\,{\sc i} line
at 6572.8~\AA\ of similar strength to the Li\,{\sc i} line is reproduced (Figure~6, bottom panel)).
This is
evidence that unusual atmospheric effects are not
seriously biassing the isotopic ratio.

Our abundance estimates assume LTE prevails. Carlsson et al. (1994)
quantified non-LTE effects for Li\,{\sc i} lines in red giants.
In the case of the 6707 \AA\ line, the correction for non-LTE effects
is sensitive to the strength of the line. For a star of $\log g \simeq
2.0$ and $T_{\rm eff} \simeq$ 4500 K with a 6707 \AA\ of $W_\lambda \simeq
400$ m\AA,  the non-LTE abundance is only about 0.1 dex lower than the
LTE abundance. The non-LTE abundance for the 6104 \AA\ line under the
same conditions is about 0.2 dex greater than the LTE result. These
adjustments are barely larger than the uncertainties of the LTE abundances
arising from the errors in the adopted atmospheric parameters, especially the
effective temperature. 

\subsection{C, N, and O Abundances}

Our approach to finding the C,N, and O abundances was a differential one involving IRAS\,19285+0517, the
Sun, and $\alpha$\,Ser. The K giant $\alpha$\,Ser is a Li-poor giant with
a $T_{\rm {eff}}$, log $g$ and [Fe/H] very similar to that of IRAS\,19285+0517.
We analysed the same lines in all three objects and
consider the differential abundances. In this way, the effects of several
potential sources of error (e.g., the CN molecule's dissociation
error) are mitigated, if not entirely eliminated.

Spectra of $\alpha$\,Ser were obtained also with the McDonald Observatory's
2.7 m telescope and analysed using our standard procedures. Our
derived fundamental parameters (Table 3) show that, to within
the errors of measurement, it is a `carbon copy' of IRAS\,19285+0517,
apart from the great difference in lithium abundance and projected rotational velocity.

Three indicators of the carbon abundance were used: the [C\,{\sc i}]
8727 \AA\ line, the C\,{\sc I} 5380 \AA\ line, and a C$_2$ Swan system
triplet at 5135 \AA\ with lower weight given to another C$_2$
feature at 5141 \AA. 
We took atomic and molecular data around the [C\,{\sc i}] line from Gustafsson et al. (1999).
For the 5380~\AA\ line
we adopted the theoretical log $gf$ = $-$1.57 (Bi\'{e}mont et al. 1993).
For the spectrum syntheses near 5135 \AA\ (Fig~7),
we took data for C$_2$ lines
from Querci, Querci, \& Kunde (1971), and for MgH and the atomic lines from
Kurucz data base.

The derived carbon abundances are given in Table 5. For
IRAS\,19285+0517 relative to Sun, [C/H] is 0.10 (8727 \AA), 0.19 (5380 \AA) (Fig~8), and
$-$0.10 (5135 \AA), for a mean of 0.07. 
This results in a carbon abundance of [C/Fe] = $-$0.07, which is a mild carbon
deficiency as compared to [C/Fe] = 0 as the initial ratio for stars of roughly solar metallicity.
For $\alpha$ Ser, [C/H] is 0.10 (8727 \AA), 0.15 (5380 \AA), and $-$0.05
(5135 \AA) for a mean of 0.07 or [C/Fe] = 0.07, a slight
carbon enrichment relative to the initial carbon abundance.

The nitrogen abundance and $^{12}$C/$^{13}$C ratio were obtained from the CN Red System lines
at 8005~\AA\ and the mean
carbon abundance. Basic data for the CN lines were taken from de Laverny
\& Gustafsson (1998, also private communication) with a dissociation
energy D$_0$ = 7.75 eV (Lambert 1993).
For IRAS\,19285+0517, [N/H] = 0.20,
which, if [N/Fe] = 0 is appropriate for the unevolved progenitor,
corresponds to a mild nitrogen enrichment of [N/Fe] = 0.06. The enrichment
for $\alpha$ Ser is more substantial: [N/H] = 0.35 and [N/Fe] = 0.35.
Syntheses of the 8005 \AA\ region yields for IRAS\,19285+0517 the
isotopic ratio $^{12}$C/$^{13}$C = 9 $\pm$ 3 from the $^{13}$CN feature
at 8004.7~\AA\ (Fig~9). The error on the isotopic
ratio was estimated as a quadratic sum of the various uncertainties.
The uncertainty in C and N abundances of $\pm$0.1~dex leads to uncertainty of
$\pm$2, respectively, in the
$^{12}$C/$^{13}$C ratio. The rms = 0.004 of the S/N at the continuum yields
approximately an uncertainty of $\pm$1 in the carbon ratio. The quoted
uncertainties in the derived model parameters have insignificant effect on the
derived $^{12}$C/$^{13}$C ratio.
The ratio for $\alpha$ Ser was determined 
to be $^{12}$C/$^{13}$C = 10$\pm3$, which is in good agreement with the ratio $^{12}$C/$^{13}$C = 12$\pm2$
previously determined by Day, Lambert, \& Sneden (1973)

Oxygen abundances are obtained from the [O\,{\sc i}] 6300 \AA\ line.
Allowance was made for a blending Ni\,{\sc i} line (Allende Prieto et al.
2001). IRAS 19285+0517 has [O/H] of 0.09, a value very close to the
likely initial value. Oxygen in $\alpha$ Ser is apparently enhanced with
[O/H] = 0.19.
A summary of the derived mean C, N, and O abundances, and the ratios relative
to Fe ([X/Fe]) for IRAS\,19285+0517, $\alpha$ Ser, and Sun is
given in Table~6.

\section{DISCUSSION}

\subsection{Defining the Anomalies}

According to the inferred luminosity and the derived effective
temperature, IRAS 19285+0517 is probably a He-core burning (clump)
giant. Since the red giant branch for H-shell burning stars at the
luminosity of the clump is
only slightly cooler than the clump, we cannot exclude the
possibility that IRAS 19285+0517 is evolving to He-core ignition.

A normal giant, whether on the RGB prior to He-core ignition or at the clump
as a  He-core burning star, has a deep convective envelope that developed
when the star was  a subgiant. The result of convection is to dilute the surface
lithium abundance by a factor of about 50 so that a normal lithium
abundance does not exceed $\log\epsilon$(Li) = 1.2. The convective
envelope also brought to the surface material exposed mildly to the
H-burning CN-cycle. This process reduces the surface $^{12}$C abundance,
reduces the $^{12}$C/$^{13}$C ratio, increases the $^{14}$N abundance,
but is predicted to leave the $^{16}$O abundance unaffected. Analyses
of red giants
confirm the sense of these changes but there is ample
evidence that observed changes, especially
the $^{12}$C/$^{13}$C ratio, are frequently more extreme than predicted
by standard models.

Our reference giant $\alpha$ Ser is a typical giant with its $^{12}$C/$^{13}$C
ratio (12$\pm3$) less than predicted by standard models. Observations of the
isotopic ratio in giants of the open cluster M\,67 show that `low' ($\leq 10$)
$^{12}$C/$^{13}$C  ratios are found only in giants that have evolved to or
beyond the tip of the red giant branch (Gilroy \& Brown 1991). If, as is
possible, $\alpha$\,Ser is a clump giant, its low isotopic ratio fits
the pattern established by M\,67. The same claim may be made for
IRAS 19285+0517 but its status as a `normal' giant is, of course,
questionable.

$\alpha$ Ser was included by Lambert \& Ries (1981) in their C, N, and O
analyses of a sample of field K giants. The expected changes in C and N
were revealed by this analysis. Relative to the mean abundances for the
sample, $\alpha$\,Ser
was somewhat less depleted in C and less enriched in N with an
approximately normal O abundance (Table~6). IRAS\, 19285+0517 is a close but not
exact match to $\alpha$\,Ser. The closeness of the match depends on
whether the comparison is made using [X/H] or [X/Fe]. Using [X/H],
the C and N abundances are very similar, possibly identical to
within the errors of measurement; $\alpha$\,Ser is slightly enriched in
N and O, where the latter is presumably a reflection of a higher initial
abundance. If [X/Fe] is the preferred indicator, IRAS\,19285+0517 is
underabundant in C, N, and O by $-$0.14, $-$0.29, and $-$0.24 dex, respectively,
relative to $\alpha$\,Ser. Judged with respect to the assumption that
initial abundances would have satisfied the condition [C/Fe] = [N/Fe] = 0,
IRAS 19285+0517 is possibly just deficient in carbon ($-$0.07 dex) and
enriched in nitrogen (+0.06 dex). Given that a typical red giant
displays larger abundance changes - [C/Fe] $\simeq$ $-$0.2 and [N/Fe]
$\simeq $ +0.4 are typical (Lambert \& Ries 1981) - there is a suspicion
that Li-rich IRAS 19285+0517 has `anomalous' carbon and nitrogen
abundances.

The certain and pronounced anomalies of IRAS\,19285+0517 are (i) its
high lithium abundance ($\log\epsilon$(Li) = 2.5),
(ii) its rapid rotation ($v\sin i = 9$ km s$^{-1}$), and
(iii) its substantial infrared excess.  In having these anomalies,
IRAS 19285+0517 is similar HD\,233517 (Balachandran et al. 2000),
and PDS\,365 (Drake et al. 2001). 
Drake et al. list the known K giants
with $v\sin i \geq 8$ km s$^{-1}$. Of their 22 stars, 6 of the 14 with a
measured lithium abundance are Li-rich. Among the few K giants with
a pronounced
infrared excess, about half are Li-rich (de la Reza et al. 1997). The
impression given is that Li enrichment, rapid rotation, and infrared
excess are causally related. That the correlation between the three
observational attributes is not perfect may be due to different
times and time scales for their  appearance and/or  disappearance; for example, surface
lithium abundance may decay as lithium is destroyed internally
at a rate different from the rate at which angular momentum is
removed from the envelope.

\subsection{Explaining the Anomalies}

An assumption common to the proposed explanations of the Li-rich
giants is that, as a subgiant, the
star underwent the predicted severe dilution of its original
lithium and the lithium was subsequently replenished  at some point in
the giant's evolution. 
This is a plausible scenario since
several Li-rich giants, some slowly rotating and others rapidly rotating,
have a lithium abundance greater than the star's presumed initial
abundance. In addition, there are no known main sequence stars
with a lithium abundance that, after standard dilution by the factor
of about 30 to 50, would provide the high lithium abundance of the Li-rich
giants. Therefore, proposed explanations envisage either lithium
production by the giant or replenishment of surface lithium by
capture of one or more planets. Explanations should be judged for
plausibility not only for the lithium
abundance but also by their ability to account for the existence of circumstellar material as revealed
through the infrared excess and for the rapid rotation of many Li-rich
giants.

Proposals invoking lithium production exploit the $^3$He reservoir
predicted to exist outside the He-core and, if present,
the H-burning shell of the red giant. This reservoir was built up
in the main sequence star outside its H-burning core in layers
sufficiently hot for $^3$He to be produced by the initial steps
of the $pp$-chain but not so hot that $^3$He was quickly destroyed.
The giant's convective envelope that diluted the surface lithium also
smeared the $^3$He reservoir over the envelope. Given this supply of
$^3$He, the reactions $^3$He$(\alpha,\gamma)^7$Be$(e,\nu)^7$Li may convert
$^3$He to $^7$Li. To achieve efficient lithium production, the $^7$Be
must be swept quickly to lower temperatures in order to avoid
destruction of $^{7}$Be and $^{7}$Li by hot protons. This scenario of
lithium production, which is commonly called the
Cameron-Fowler (1971) mechanism, accounts for the Li-rich AGB stars
which are predicted to have a deep convective envelope with a hot bottom.

The $^3$He reservoir has been invoked by several authors (e.g., Sackmann \& Boothroyd 1999) as the key to the Li-rich
giants (i.e stars at luminosities too low for them to be identified as AGB giants) but, in many
cases, the physical trigger for $^7$Li production has not been
identified.
Palacios, Charbonnel, \& Forestini (2001) have suggested
that the burning of $^7$Li created by depletion of the $^3$He
reservoir enables lithium to get to the surface. Rapid rotation
is suggested as a way to generate a `lithium flash' to enhance the
mixing and surface abundance. This process is expected to occur
when the red giant's H-burning shell burns through a discontinuity in
mean molecular weight introduced previously by the development of the
convective envelope. A majority of the Li-rich giants appear to
be at the predicted luminosity (Charbonnel \& Balachandran 2000),
as estimated from their {\it Hipparcos} parallaxes. (The luminosity
and $T_{\rm eff}$ are not very different from the prediction
for He-core burning giants.) This interesting idea only in a more
speculative fashion accounts for the infrared excess; the luminosity
increase of about a factor of 2 during the brief flash
(duration $\sim$ 10$^3$ yrs) is blamed for the mass loss on the
empirical evidence from much more luminous stars that mass loss
increases with luminosity. The idea does explain why lithium is
pure $^7$Li. If our suspicion is confirmed that the carbon and nitrogen abundances
of IRAS 19285+0517 are less extreme than expected for a normal red giant,
this would seem to present a difficulty for the Li-flash
scenario; the predicted duration of the flash is too short to affect the
abundances which should reflect fully the alterations induced by the
convective envelope.

An alternative explanation for the main anomalies (Li enrichment,
rapid rotation, and circumstellar matter) is the accretion  by
the red giant of material
in the form of planets or a brown dwarf. This has been discussed to
different degrees of detail by Alexander (1967), Siess \& Livio (1999) and
and Denissenkov \& Weiss (2000).

Siess \& Livio (1999) discuss the accretion of a planet or brown
dwarf by 1~M$_{\odot}$ red giants of different luminosities.
It is shown that the red giant may be
spun up, eject mass, and become enriched in lithium. Considerable adjustments
to the  internal structure occur but the surface lithium enrichment results
entirely from the lithium in the accreted material. 
If destruction of lithium is avoided, the maximum lithium
abundance resulting from the accretion of giant planets or a brown
dwarf is necessarily the interstellar value (say, $\log\epsilon$(Li)
$ \simeq$ 3). The lithium abundance of IRAS\,19285+0517 is comfortably below
this limit but several Li-rich stars (Charbonnel \& Balachandran 2000;
Drake et al. 2001) have lithium above this limit. In principle, higher
lithium abundances are possible through the accretion of terrestrial
planets (i.e., cosmic material less hydrogen, helium, and other volatiles),
but this requires that the mass of the material from which the terrestrial
planets were derived exceeds the mass of the red giant (Brown et al. 1989),
which would seem to be a difficult condition to meet.
Denissenkov \& Weiss (2000) propose that accretion activates the
Cameron-Fowler mechanism and thus the surface lithium abundance
may exceed the initial value before destruction sets in.

In addition to accounting for the lithium, the rapid rotation, and the
infrared excess, accretion of giant planets or a brown dwarf 
serves to adjust the C and N abundances towards their initial values and away
from the changed values induced by the convective envelope. 
If the lithium
is added without subsequent destruction, the isotopic lithium
ratio would be expected to be close to the initial value for the
star and its natal cloud, which, on the basis of limited evidence
from measurements of the ratio in local diffuse clouds, is likely to
have been about $^7$Li/$^6$Li = 10, a ratio clearly ruled out
for IRAS 19285+0517. However, for this star, we cannot
exclude the possibility that a small amount of
lithium was destroyed subsequent to the
accretion. Since $^6$Li is destroyed about 70 times faster than $^7$Li,
even the loss of trace amounts of lithium ensures effectively complete
destruction of $^6$Li.

\section{CONCLUDING REMARKS}

With the discovery of extra solar planets in close elliptical orbits, accretion of giant planets
by red giants is a virtual certainty. That this phenomenon can
account for all of the known Li-rich giants is uncertain. 
Accretion without activation of the $^{3}$He reservoir certainly cannot account for
the super Li-rich stars which have more Li than the ISM value.
It would be helpful to extend the detailed
analyses of the Li-rich giants. Extensions should include a uniform
abundance analysis for the light elements (and isotopes) C, N, and O
for a sample including normal K giants, so that subtle differences
between Li-rich and Li-normal giants may be distinguished.
Extant analyses show that there are no large differences (Berdyugina \&
Savanov 1994; da Silva et al. 1995).
Measurement of the beryllium abundance in Li-rich giants would
be valuable. Accretion of a planet or brown dwarf restores the Be abundance.
Conversion of the $^{3}$He reservoir to $^{7}$Li does not return the
Be abundance to its pre-diluted value (Castilho et al. 1999).

Measurements of the $^7$Li/$^6$Li ratio should be extended to other
stars. Positive detection of $^{6}$Li would favor the accretion
scenario.
In this regard, we note the recent detection of $^6$Li in the
atmosphere of a main sequence star with a giant extra solar planet 
(Israelian et al. 2001) suggests that the
star has accreted one or more former planets. Early accretion of this kind
will not account for Li-rich giants like IRAS 19285+0517, because in evolution
to the red giant phase, the convective envelope will dilute the lithium
to below the observed abundance of IRAS\,19285+0517.

\acknowledgments
We acknowledge the support of the National Science Foundation (Grant AST-9618414) 
and the Robert A. Welch Foundation
of Houston, Texas (Grant F-634). 
BJH acknowledges support from the National Science Foundation
(AST-9315107, AST-9900846).
This research has made use
of the Simbad database, operated at CDS, Strasbourg, France, and the NASA ADS service, USA.

\clearpage

\figcaption[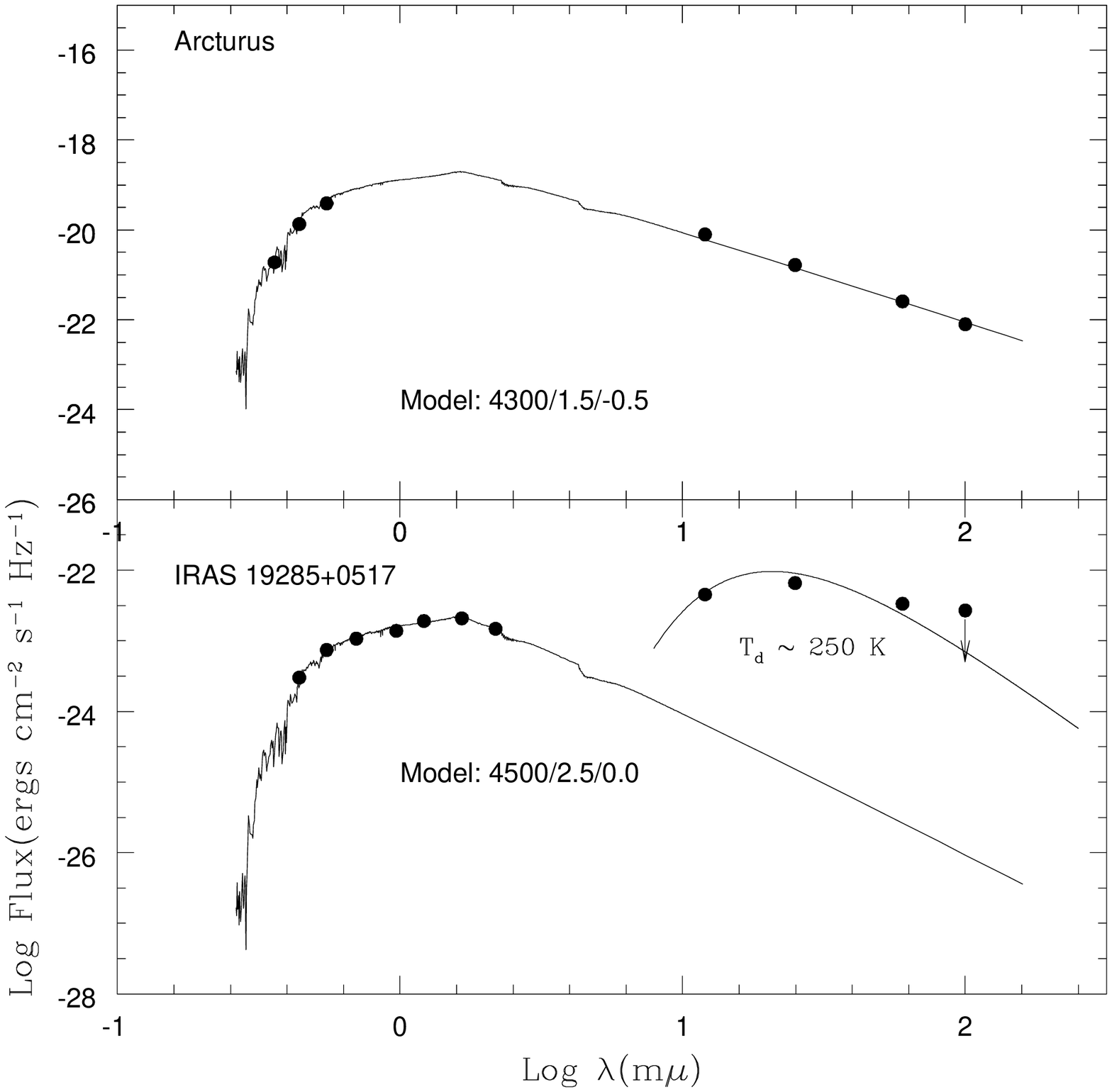] {Spectral energy distributions of Arcturus (top panel) and
IRAS\,19285+0517 (bottom panel). The solid line in both cases is the predicted
flux distribution from the Kurucz model atmosphere 
having the
fundamental parameters described in the text and given in the
figure. The infrared excess of IRAS\,19285+0517 at 12, 25, 60, and 100 $\mu$m
is evident and is fit with a 250 K black body curve \label {fig1}}

\figcaption[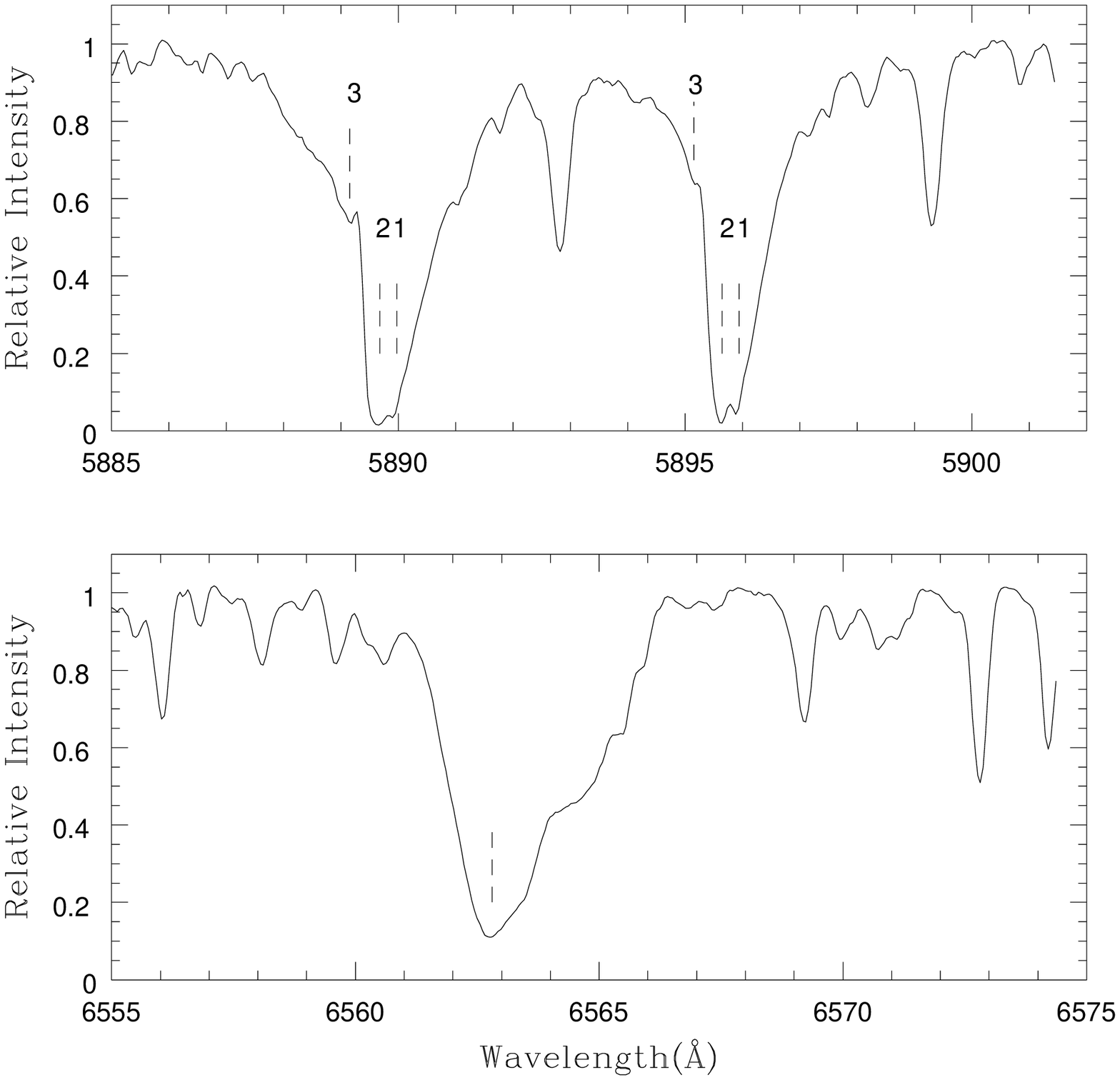]{The spectrum of IRAS\,19285+0517 around the Na\,D lines (top panel)
and the H$\alpha$ line (bottom panel). The vertical dashed lines labled as 1, 2, and 3 (top panel) indicate
photospheric velocity, $-$16 km s$^{-1}$, and $-$38 km s$^{-1}$, respectively (see the text for details).
The vertical dashed line (bottom panel)
represents the photospheric velocity of the H$\alpha$. \label {fig2}}

\figcaption[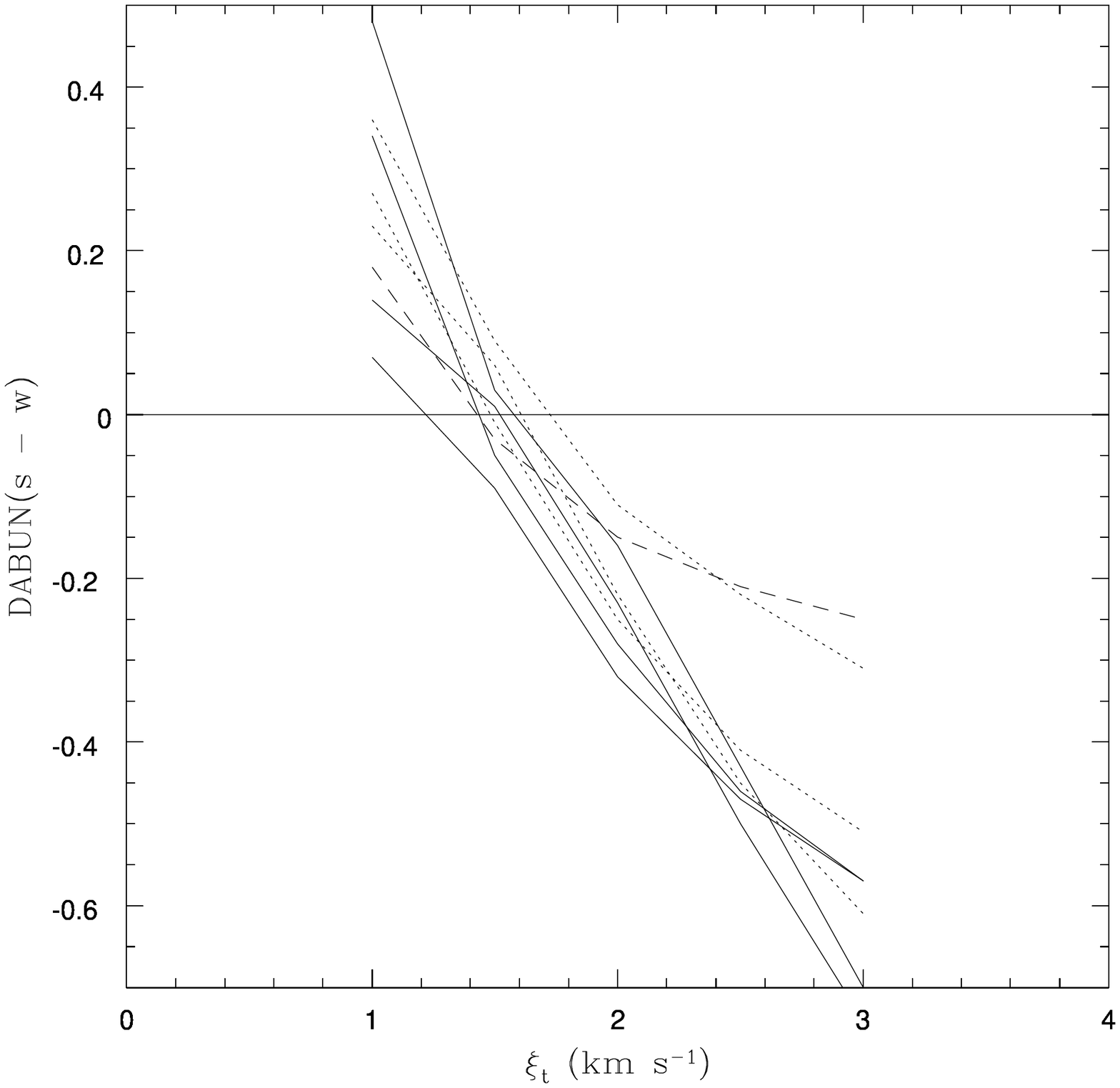] {Abundance difference between combinations of strong and weak lines
of similar lower excitation potential from the same atom, DABUN(s-w),
plotted against the assumed microturbulence. Results from lines of iron (solid lines),
cobalt (dashed line), and nickel (dotted lines) are shown. A microturbulence
$\xi_{\rm t}$ = 1.5 $\pm$ 0.1 km s$^{-1}$ corresponding to
DABUN(s-w) = 0 is adopted. \label{fig3}}

\figcaption[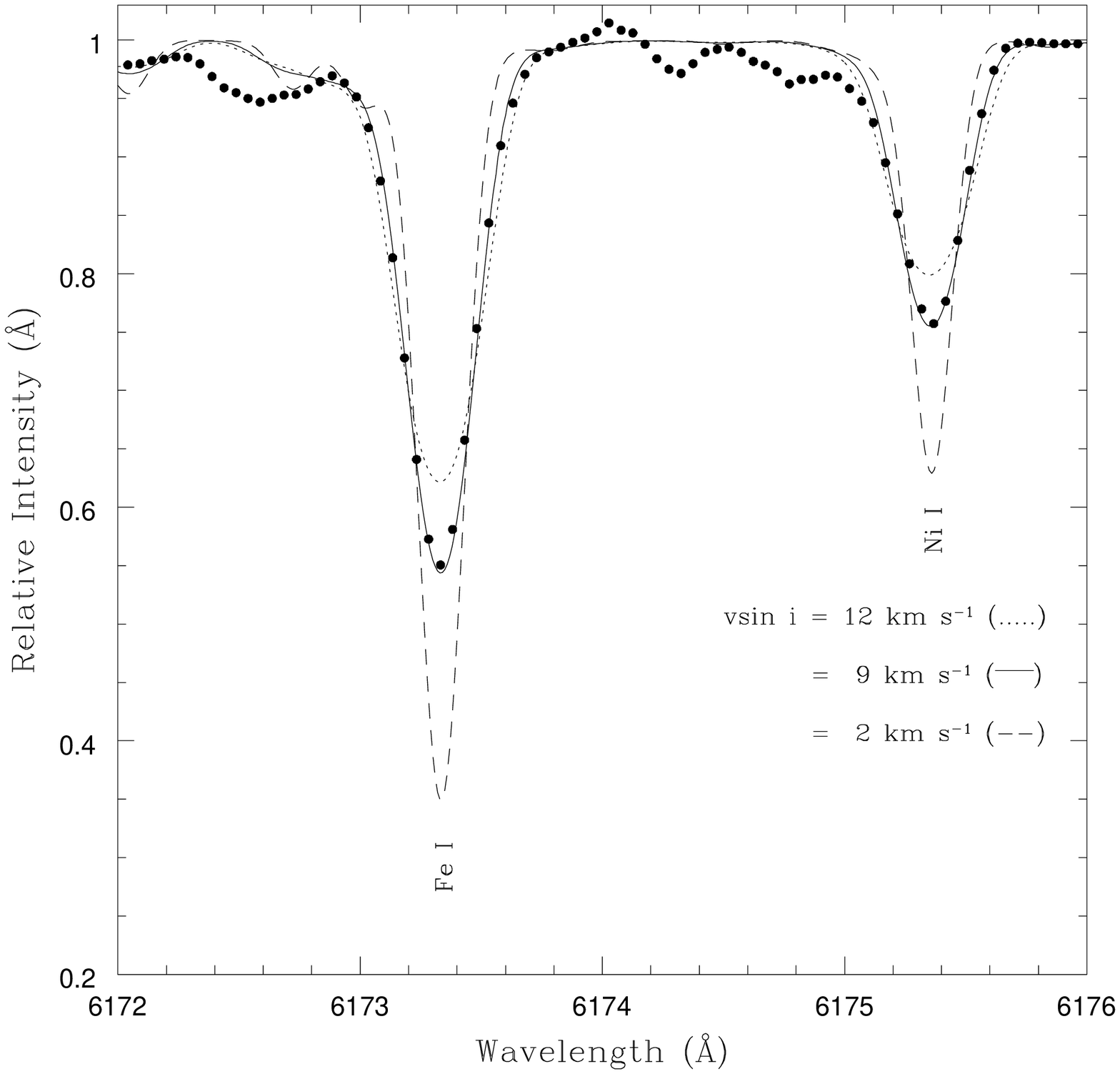]{The spectrum of IRAS\,19285+0517 from 6172 \AA\ to 6176 \AA\ showing
strong Fe\,{\sc i} and Ni\,{\sc i} lines. The observed spectrum (filled
circles) is shown together with three predicted line profiles
computed from the adopted model atmosphere and identical assumptions
but for the  three different values of the projected rotational
velocity: $v\sin i = 2, 9$, and 12 km s$^{-1}$. The velocity
$v\sin i = 9$ km s$^{-1}$ provides a good fit to the observed line
profiles. See text for additional details. \label{fig4}}

\figcaption[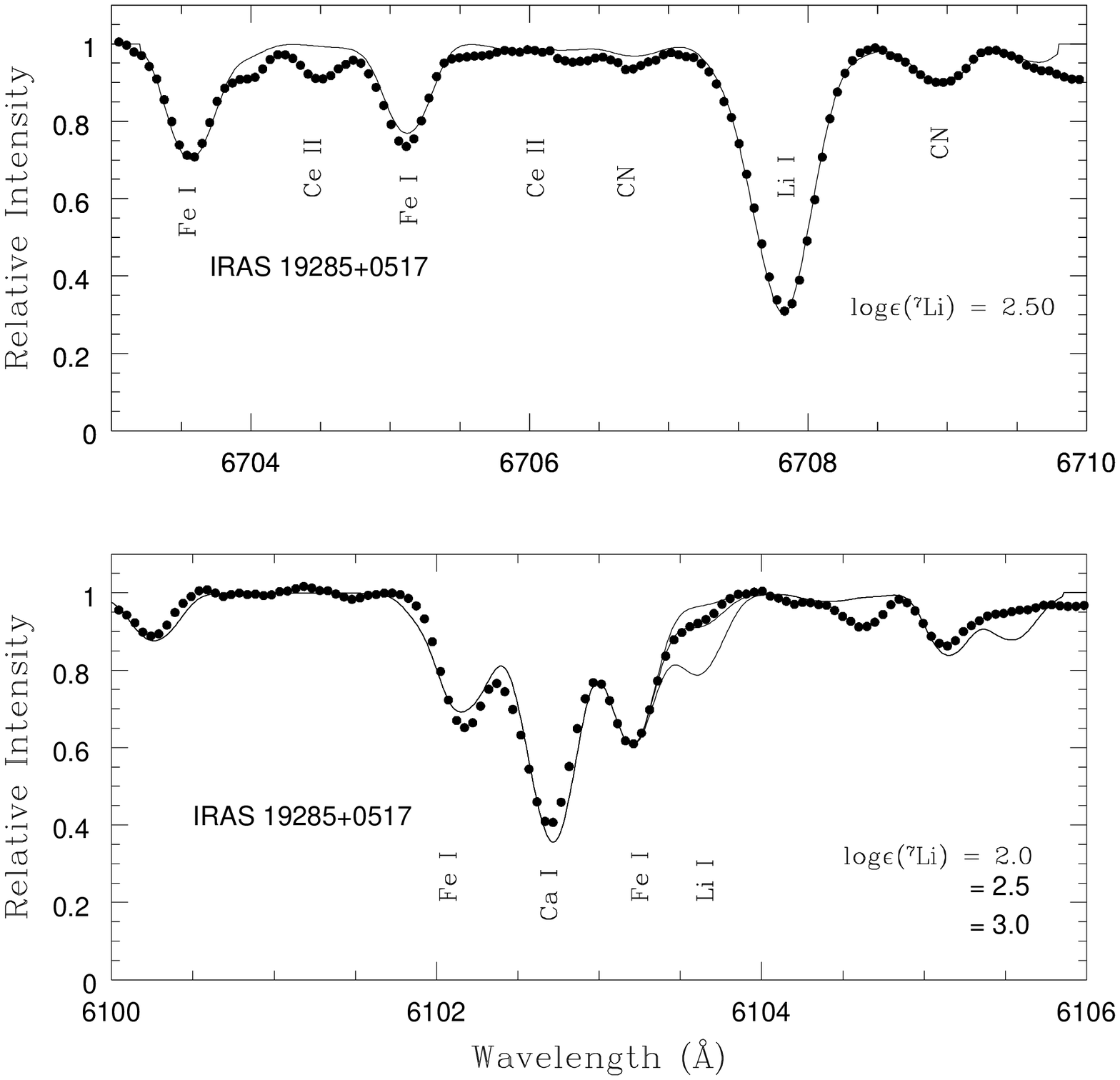]{Observed (filled circles) and synthetic spectra around the Li\,{\sc i}
lines at 6707 \AA\ (top panel) and 6104 \AA\ (bottom panel).
The observed 6707 \AA\ line is well matched by the  synthetic spectrum
computed for the abundance $\log\epsilon$($^7$Li) = 2.50. Three synthetic
spectrum are shown for the 6104 \AA\ line, with that for
$\log\epsilon$($^7$Li) = 2.5 providing a satisfactory fit to the
observed spectrum. \label{fig5}}

\figcaption[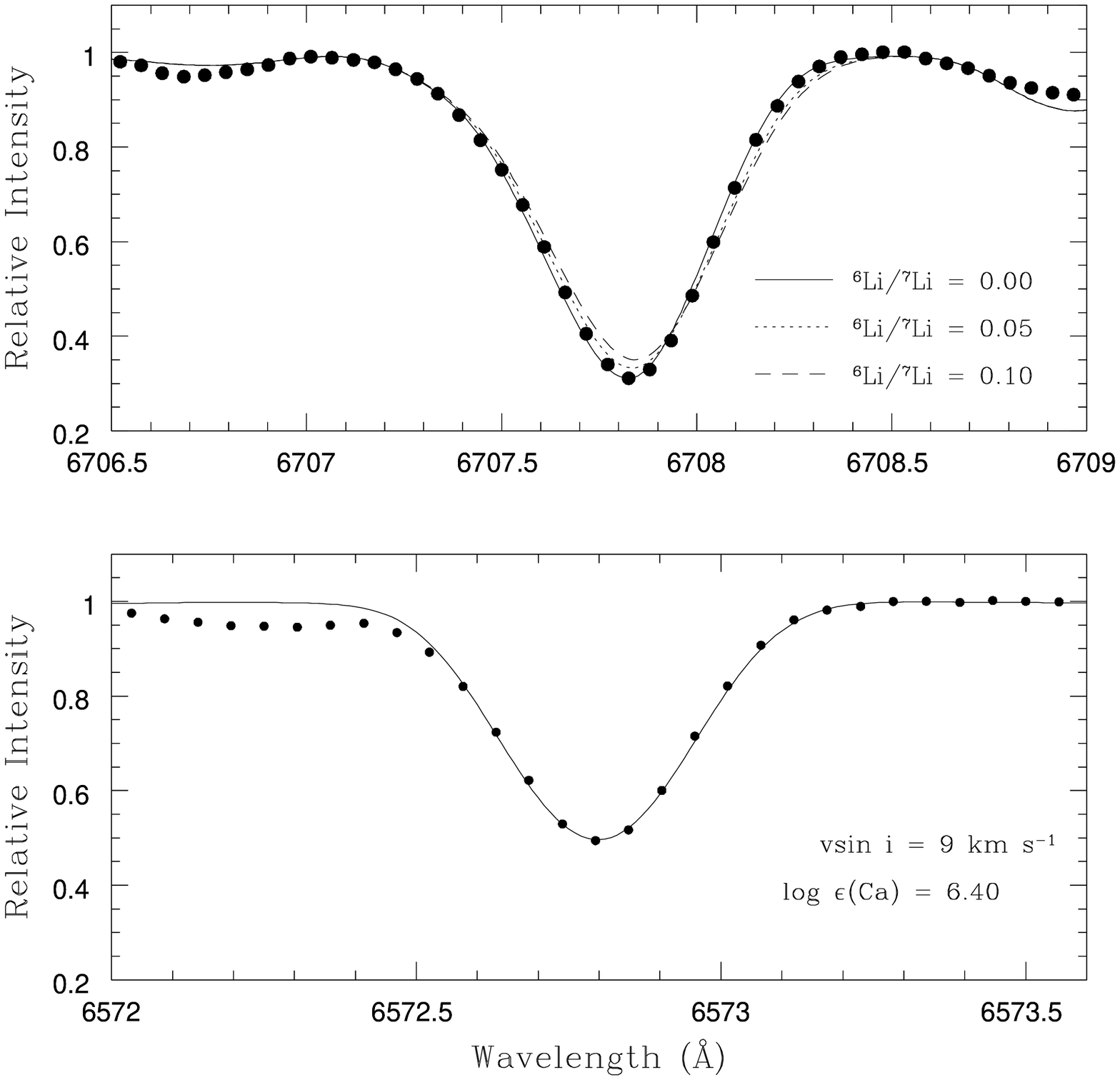]{Observed (filled circles) and synthetic spectra of the Li\,{\sc i} 6707 \AA\
resonance doublet (top panel) and the Ca\,{\sc i} intercombination
resonance line at 6572 \AA\ (bottom panel).
In the top panel, synthetic spectra are provided for the three
isotopic ratios $^6$Li/$^7$Li = 0.0, 0.05, and 0.10, with $^{6}$Li/$^{7}$Li = 0.0 providing
the best fit. The synthetic
spectrum in the bottom panel shows that observed Ca\,{\sc i} line
is well matched by the synthetic line. \label{fig6}}

\figcaption[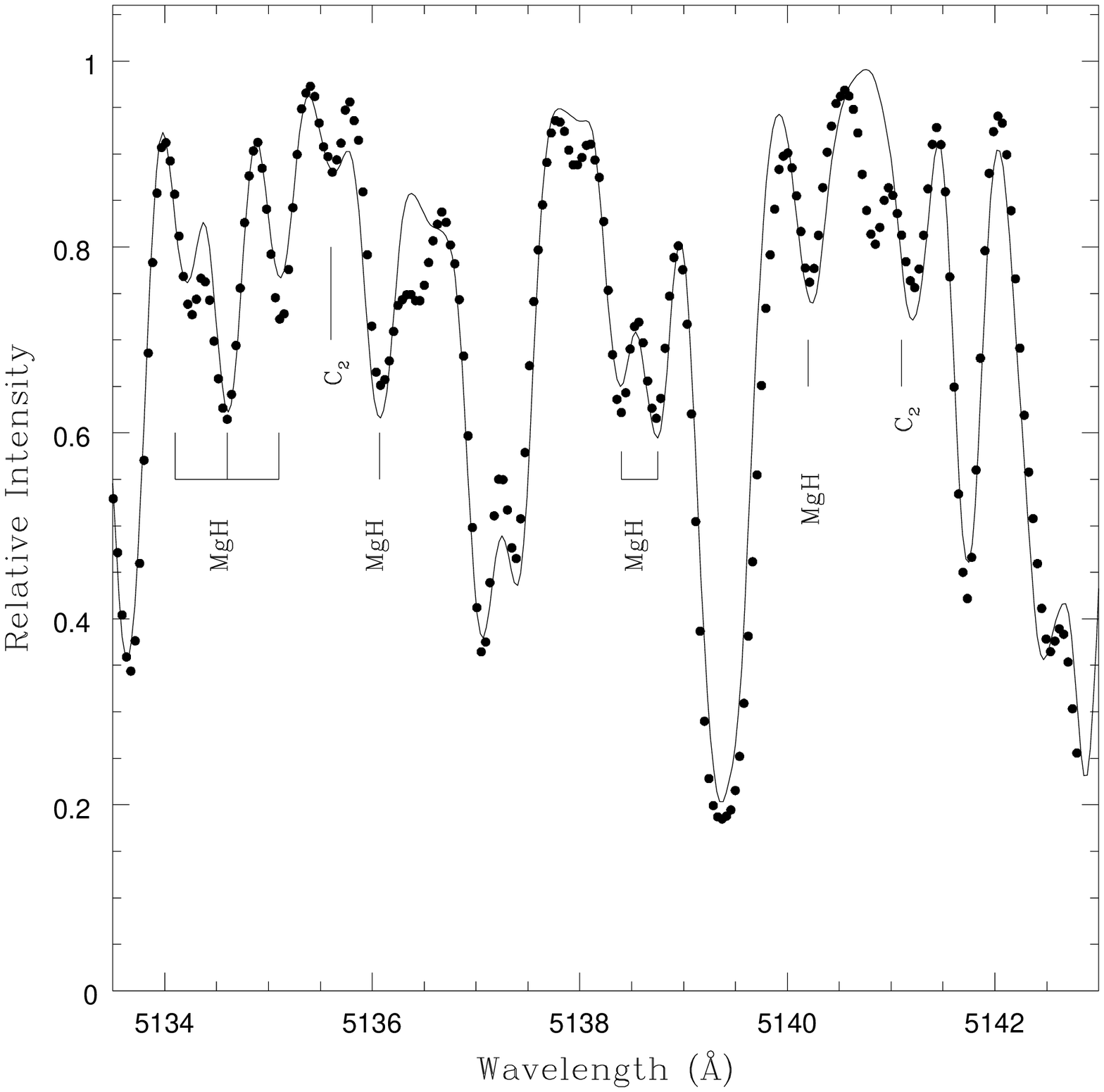]{The spectrum of IRAS\,19285+0517 around 5138 \AA. Several MgH and C$_2$ lines
are identified including the C$_2$ feature at 5135.6 \AA, a primary
indicator of the carbon abundance. \label{fig7}}

\figcaption[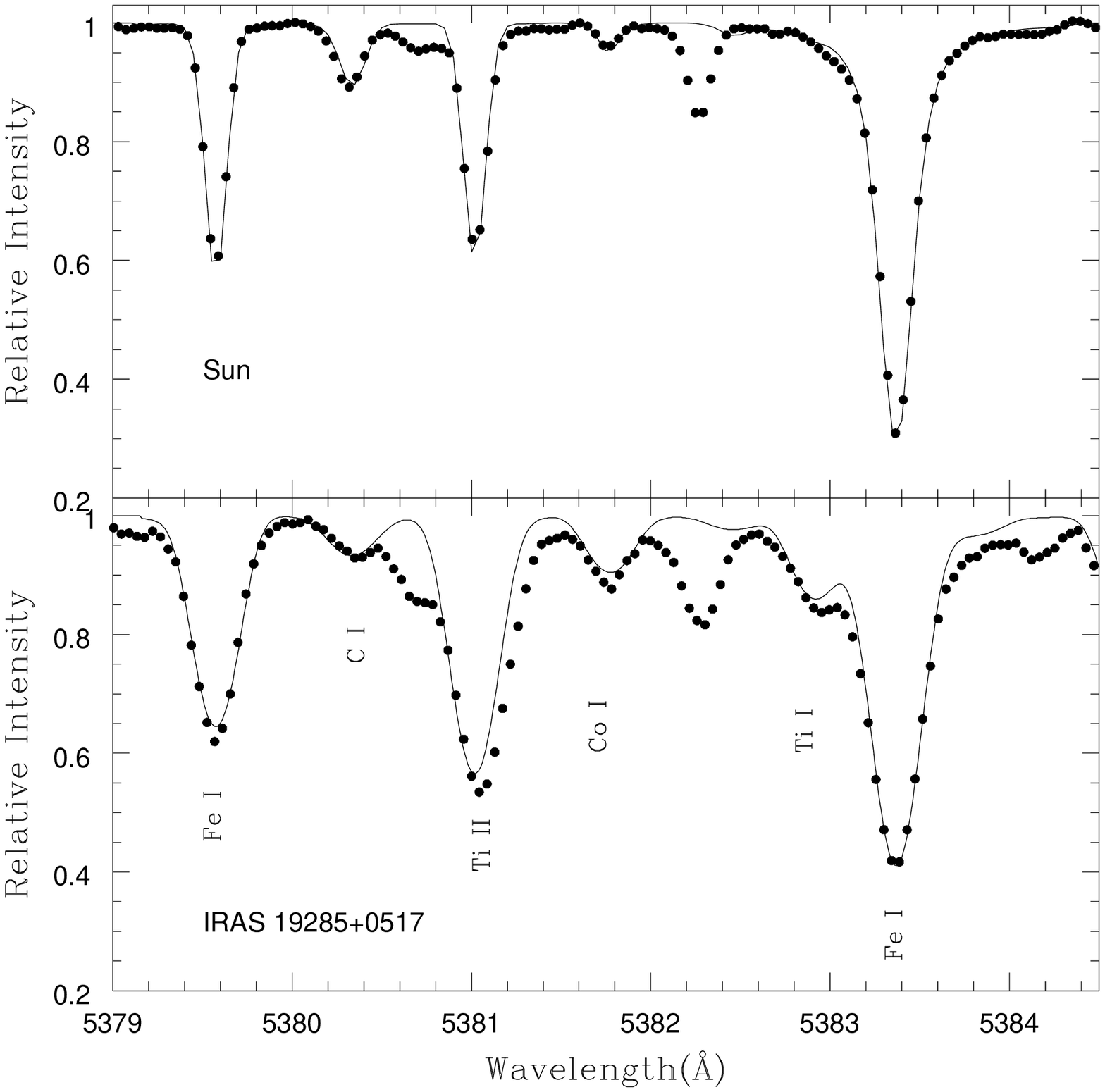]{The spectrum of the Sun (top panel)
and IRAS\,19285+0517 (bottom panel) from 5379 \AA\ to 5384 \AA\
showing the C\,{\sc i} the line at 5380 \AA. The synthetic spectrum (solid line) shown
for IRAS\,19285+0517 is computed for the carbon abundance $\log\epsilon$(C) =
8.66. \label{fig8}}

\figcaption[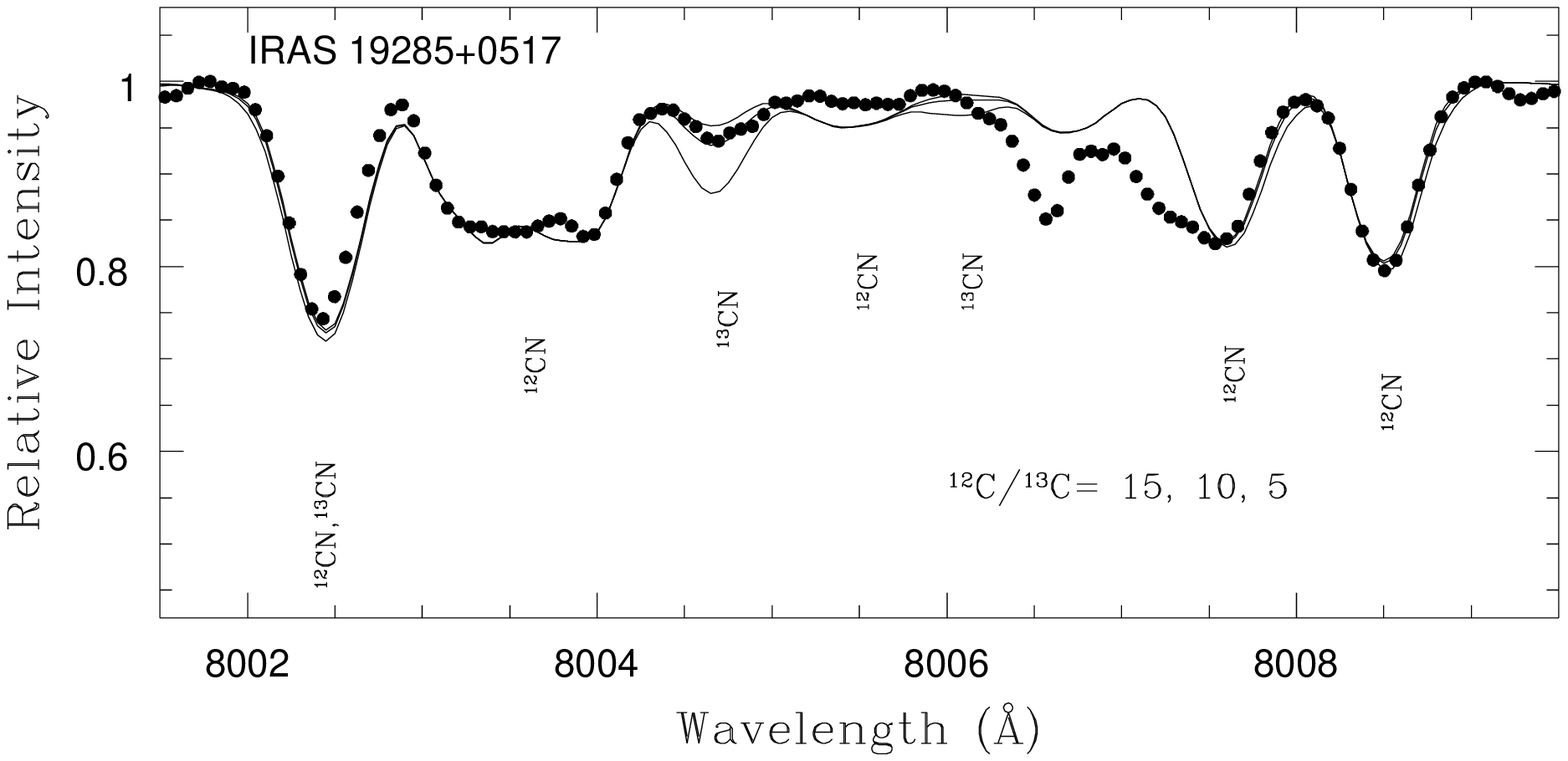]{The spectrum of IRAS\,19285+0517 around 8005 \AA. Synthetic spectra (solid line)
show the contributions of CN Red system lines with the $^{12}$C and
$^{14}$N  abundances fixed but the $^{12}$C/$^{13}$C ratio set
at 5, 10, and 15. \label{fig9}}

\clearpage
\renewcommand{\arraystretch}{0.8}
\begin{table}
\caption{Ground-Based Photometry for IRAS~19285+0517}
\begin{tabular}{ccc}
\tableline
\tableline
Band   & Mag. & log (F$_{\nu}$)$^{a}$ \\
       &      & (ergs cm$^{-2}$ s$^{-1}$ Hz$^{-1}$) \\
\tableline
B      & 11.83$\pm0.01$ &  $-$23.52 \\
V      & 10.35$\pm0.02$ &  $-$23.13 \\
R$_{c}$& 9.52$\pm0.03$  &  $-$22.97 \\
I$_{c}$& 8.78$\pm0.03$  &  $-$22.86 \\
J      & 7.71$\pm0.01$  &  $-$22.72\\
H      & 6.98$\pm0.02$  &  $-$22.68 \\
K      & 6.70$\pm0.01$  &  $-$22.83 \\   
\tableline
\end{tabular}
\tablenotetext{a}{fluxes are corrected for a reddening of E(B-V) = 0.34.}
\end{table}

\clearpage

\begin{table}
\caption{Lines used in Deriving $\xi_{t}$}
\begin{tabular}{ccccc}
\tableline
\tableline
Line      &   Iden.   &  LEP    &   log $gf$&         W$_{\lambda}$ \\
(\AA)     &           & (eV)    &           &        (m\AA) \\ 
\tableline
6692.265  &  Fe\,{\sc i}   &  4.076  &  -2.899   &                 21.3  \\
6027.050  & Fe\,{\sc i}   &  4.076  &  -1.090   &                 106.0 \\
6699.200  & Fe\,{\sc i}   &  4.590  &  -2.191   &                 41.0\\
6024.049  & Fe\,{\sc i}   &  4.548  &  -0.060   &                 165.0\\
5367.470  & Fe\,{\sc i}   &  4.415  &   0.350   &                 182.3\\
6704.500  & Fe\,{\sc i}   &  4.220  &  -2.660   &                 31.0\\
6159.368  & Fe\,{\sc i}   &  4.607  &  -1.970   &                 52.0\\
6293.923  & Fe\,{\sc i}   &  4.835  &  -1.605   &                 50.0\\
6055.992  & Fe\,{\sc i}   &  4.733  &  -0.460   &                 104.0\\
5987.066  & Fe\,{\sc i}   &  4.795  &  -0.426   &                 115.0\\
6005.029  & Co\,{\sc i}   &  1.711  &  -3.320   &                 30.0\\
6116.990  & Co\,{\sc i}   &  1.785  &  -2.490   &                 72.0\\
6093.140  & Co\,{\sc i}   &  1.740  &  -2.440   &                 77.0\\
7062.970  & Ni\,{\sc i}   &  1.951  &  -3.500   &                 75.0\\
6914.560  & Ni\,{\sc i}   &  1.951  &  -2.070   &                 158.0\\
6025.730  & Ni\,{\sc i}   &  4.236  &  -1.760   &                 18.0\\
6086.290  & Ni\,{\sc i}   &  4.266  &  -0.530   &                 85.0\\
6133.950  & Ni\,{\sc i}   &  4.088  &  -1.830   &                 31.0\\
6176.810  & Ni\,{\sc i}   &  4.088  &  -0.530   &                 102.0\\
\tableline
\end{tabular}
\end{table}

\clearpage

\begin{table}
\caption{Derived Fundamental Properties}
\begin{tabular}{ccccc}
\tableline
\tableline
Star&\lbrack Fe/H\rbrack & $T_{\rm {eff}}$  & log $g$  & $\xi_{t}$  \\
    &                    & (K) & (cm s$^{-2}$) & (km s$^{-1}$) \\
\tableline
IRAS\,19285+0517&0.14$\pm0.16$& 4500$\pm$100  & 2.50$\pm$0.25 & 1.55$\pm$0.1  \\
Arcturus & $-$0.58$\pm$0.1 & 4275$\pm$50 & 1.5$\pm$0.25 & 1.6$\pm$0.3  \\
         &   ($-$0.50$\pm$0.1)$^{a}$ & (4300$\pm$30)$^{a}$ & (1.5$\pm$0.15)$^{a}$ & (1.7)$^{a}$ \\
$\alpha$ Ser & 0.0$\pm$0.1 & 4600$\pm$100 & 2.50$\pm$0.25 & 1.9$\pm$0.25 \\
\tableline
\end{tabular}
\tablenotetext{a}{from Peterson, Dalle Ore, \& Kurucz 1993.}
\end{table}

\clearpage

\begin{table}
\caption{Elemental Abundances of IRAS~19285+0517}
\begin{tabular}{cccccrr}
\tableline
\tableline
Element & log $\epsilon$(X)$_{\odot}$$^{a}$ & $n$ & log $\epsilon$(X) & $\sigma$$_{tot}$ & \lbrack X/H \rbrack & \lbrack X/Fe \rbrack \\
\tableline
Li\,{\sc i}     &1.10 & 2 & 2.50 & 0.20   &    ...  & ... \\ 
Al\,{\sc i}    & 6.47 & 3 & 6.73 & 0.07   &    0.26 &    0.12 \\
Si\,{\sc i}    & 7.55 &14 & 7.83 & 0.08   &    0.28 &    0.14 \\
Ca\,{\sc i}    & 6.36 & 3 & 6.30 & 0.20   & $-$0.08 & $-$0.22 \\
Ti\,{\sc i}    & 5.02 & 9 & 5.08 & 0.21   &    0.06 & $-$0.08 \\
Ti\,{\sc ii}   & ...  & 4 & 5.04 & 0.23   &    0.02 & $-$0.12 \\
V\,{\sc i}     & 4.00 &10 & 4.35 & 0.20   &    0.35 &   0.21 \\
Cr\,{\sc i}    & 5.67 & 7 & 5.71 & 0.14   &    0.04 & $-$0.10 \\
Cr\,{\sc ii}    & ... & 2 & 6.07 & 0.12   &    0.40 &  0.26 \\
Fe\,{\sc i}    & 7.50 &40 & 7.65 & 0.16   &    0.15 & 0.01 \\
Fe\,{\sc ii}   & ...  & 8 & 7.62 & 0.20   &    0.12 & $-$0.02 \\
Co\,{\sc i}    & 4.92 & 9 & 5.28 & 0.17   &    0.36 & 0.22 \\
Ni\,{\sc i}    & 6.25 &35 & 6.55 & 0.17   &    0.30 &  0.16 \\ 
Zr\,{\sc i}    & 2.60 & 8 & 2.66 & 0.19   &    0.06 & $-$0.08\\
Ba\,{\sc ii}   & 2.13 & 3 & 2.67 & 0.22   &    0.54 & 0.40 \\
La\,{\sc ii}   & 1.17 & 3 & 1.54 & 0.22   &    0.37 & 0.23 \\
Nd\,{\sc ii}   & 1.50 & 5 & 1.72 & 0.20   &    0.22 & 0.08 \\
\tableline
\end{tabular}
\tablenotetext{a}{from Grevesse \& Sauval 1998.}
\end{table}
\clearpage
\begin{table}
\caption{C, N, and O Abundance Analysis of IRAS~19285+0517, the Sun, and
$\alpha$ Ser}
\begin{tabular}{lccc}
\tableline
\tableline
        &    Sun    & $\alpha$ Ser  & IRAS~19285+0517 \\
Feature & log $\epsilon$(X) & log $\epsilon$(X)  & log $\epsilon$(X)  \\
\tableline
{\underline {Carbon}}                 &          &       &    \\ 
C$_{2}$ 5135~\AA\ & 8.60     & 8.55   &  8.50  \\
C\,{\sc i} 5380~\AA\ & 8.51  & 8.66 &  8.70 \\
\lbrack C\,{\sc i}\rbrack\, 8727~\AA\ & 8.50 & 8.60 & 8.60 \\
{\underline {Nitrogen}}                 &                 &    \\ 
CN (N) 8005\AA\  &  8.00 & 8.35 & 8.20  \\
{\underline {Oxygen}}     &                      &           \\ 
\lbrack O\,{\sc i}\rbrack 6300~\AA\ & 8.75 & 8.94 &  8.84  \\ 
\tableline
\end{tabular}
\end{table}

\clearpage

\begin{table}
\caption{C, N, and O Abundance Summary of IRAS\,19285+0517, the Sun, and
$\alpha$ Ser }
\begin{tabular}{rccccccccccc}
\tableline
Star  & \lbrack Fe/H\rbrack  & log $\epsilon$(C)& log $\epsilon$(N) &
log $\epsilon$(O) &  [C/Fe]&  [N/Fe] & [O/Fe] &  $^{12}$C/$^{13}$C \\
\tableline
\tableline
IRAS~19825+0517 & 0.14 & 8.60 & 8.20 & 8.84& $-$0.07 & 0.06 & $-$0.05 & 9$\pm1$ \\
$\alpha$ Ser & 0.0  & 8.60 & 8.35 & 8.94 & 0.07 & 0.35 & 0.19  & 10$\pm1$ \\ 
Sun & 0.0  & 8.54 & 8.00 & 8.75 & 0.0 & 0.0 & 0.0 & 90 \\
\tableline
\end{tabular}
\end{table}

\clearpage

\begin{figure}
\plotone{reddy.f1.eps}
\end{figure}
\clearpage

\begin{figure}
\plotone{reddy.f2.eps}
\end{figure}

\clearpage

\begin{figure}
\plotone{reddy.f3.eps}
\end{figure}

\clearpage

\begin{figure}
\plotone{reddy.f4.eps}
\end{figure}

\clearpage
\begin{figure}
\plotone{reddy.f5.eps}
\end{figure}

\clearpage
\begin{figure}
\plotone{reddy.f6.eps}
\end{figure}

\clearpage
\begin{figure}
\plotone{reddy.f7.eps}
\end{figure}

\clearpage
\begin{figure}
\plotone{reddy.f8.eps}
\end{figure}

\clearpage
\begin{figure}
\plotone{reddy.f9.eps}
\end{figure}

\clearpage

\end{document}